\documentstyle[preprint,aps]{revtex}
\tighten
\begin{document}
        
\title{\bf    Resonant continuum in the Hartree-Fock+BCS approximation}

\author{\rm N. Sandulescu $^{1,}$\cite{byline}, 
Nguyen Van Giai$^{2}$ and R. J. Liotta$^{1}$}

\address{ 
  ${1}$~ Royal Institute of Technology, Frescativ. 24,
                 S-10405 Stockholm, Sweden \\
  ${2}$~ Institut de Physique Nucl\'eaire, Universit\'e Paris-Sud, F-91406 
              Orsay Cedex, France }

\maketitle

\begin{abstract}

 A method for incorporating the effect of the resonant continuum into
 Hartree-Fock+BCS equations is proposed. The method is applied for the
 case of a neutron-rich nucleus calculated with a 
 Skyrme-type force plus a zero-range pairing interaction and the results are 
 compared with Hartree-Fock-Bogoliubov calculations.   
 It is shown that the widths of resonant states have an important 
 effect on the pairing properties of nuclei close to the drip line.

\end{abstract}
 

 In nuclei far from the $\beta$-stability line there are circumstances in 
 which the resonant part of the particle continuum  plays an important 
 role. One of the first examples was given by Migdal \cite{Migdal}, who
 anticipated  the  existence of  halo nuclei  like  $^{11}$Li  by 
 showing that two neutrons in a potential well can form a dineutron state
 in the region of the nuclear surface if the potential has a
 resonant level with energy close to zero. Here, the role of the resonance is
 to enhance the pairing correlations of the two neutrons whose attraction
 is otherwise  too weak to bind the system. 
 The interplay between the resonant 
 continuum and pairing correlations can also be important for the 
 estimation of unbound processes like 
 single-particle excitations lying above the particle emission threshold, 
 which may be found especially in nuclei close to the drip line. 
 Thus, if the excitation energy is close to the energy of a narrow resonance 
 then one expects an excitation of one-quasiparticle type, in which the 
 excited nucleon stays a finite time before decaying.
 Besides  the unbound processes mentioned above, resonant  states 
 are important for determining the pairing properties of
 the ground state of bound nuclei far from the $\beta$-stability line.
 Although in such calculations one should consider in principle the 
 complete particle continuum,
 the largest contribution to the pairing correlations is  expected to come
 from the resonant continuum part \cite{Bennett}.
  
 The aim of this paper is to propose a method for incorporating 
 the effect of the resonant continuum in the Hartree-Fock+BCS 
 (HF+BCS) approximation. 
 More precisely we investigate here the effect of the width of the 
 resonances on the pairing properties of nuclei far from stability. 
 As discussed below, this effect is difficult to estimate in  
 self-consistent Hartree-Fock-Bogoliubov (HFB) calculations 
 which are presently  used for describing  pairing correlations 
 in nuclei close to the drip line.

 In order to derive the BCS equations in the presence of the continuum
 we  discretize first the single-particle continuum spectrum 
 by enclosing the nucleus inside a box of a very large radius $R_b$.
 This is only a formal step, since will be seen that the parameter $R_b$ 
 does not appear in the final results.
 Thus the genuine continuum is replaced by a set of discrete states
 with the level density given by\cite{Bonche}:
\begin{eqnarray}\label{eq:gc}
g(\epsilon)
 & = & \sum_{\nu}\{g_\nu(\epsilon)+g_\nu^{free}(\epsilon)\} 
 \equiv  \sum_{\nu} \tilde{g}_\nu~,
\end{eqnarray}
where $g_\nu(\epsilon) \equiv (1/\pi)(2 j_\nu +1)(d\delta_{\nu}/d\epsilon)$
is the so-called continuum level density\cite{Bet37} and 
$\delta_\nu$ is the phase shift of angular momentum 
$\nu \equiv (l_\nu,j_\nu)$. 
The quantity $g_\nu^{free}(\epsilon)$ is the level density  
in the absence of the mean field and is given by
$g_\nu^{free}(\epsilon) \equiv (1/\pi)(2 j_\nu +1)(dk/d\epsilon)R_b$,
where $k$ is the momentum corresponding to the energy $\epsilon$. 
The wave functions corresponding to the positive energy states and normalized
within the box are defined  by 
  $\psi_\nu(\epsilon,r) \equiv {\it N}^{-1/2}_\nu(\epsilon) 
   \varphi_\nu(\epsilon,r)$, 
 where ${\it N}_\nu(\epsilon)$ is the norm 
 of the scattering state $\varphi_\nu(\epsilon,r)$ in the 
 box volume. It can be easily shown that for the scattering states selected
 by the box ${\it N}_\nu(\epsilon)=(2j_\nu+1)^{-1}\tilde{g}_\nu(\epsilon)$. 

 The gap equations for the states in the box can be written in terms 
of level density as follows:
\begin{equation}\label{eq:gap1}
\Delta_i = \sum_{j}V_{i\overline{i},j\overline{j}} u_j v_j +
\sum_\nu\int_{I_\nu} \tilde{g}_{\nu}(\epsilon) 
\tilde{V}_{i\overline{i},\nu\epsilon\overline{\nu\epsilon}}
u_\nu(\epsilon) v_\nu(\epsilon) d\epsilon~,
\end{equation}
\begin{equation}\label{eq:gap2}
\Delta_\nu(\epsilon) = 
\sum_{j}
\tilde{V}_{\nu\epsilon\overline{\nu\epsilon},j\overline{j}} u_j v_j +
\sum_{\nu^\prime}\int_{I_{\nu^\prime}} \tilde{g}_{\nu^\prime}(\epsilon^\prime)
\tilde{V}_{\nu\epsilon\overline{\nu\epsilon},
\nu^\prime\epsilon^\prime\overline{\nu^\prime\epsilon^\prime}}
u_{\nu^\prime}(\epsilon^\prime) v_{\nu^\prime}(\epsilon^\prime) 
d\epsilon^\prime~,
\end{equation}
where the indices $i,j$ run over the bound states and
$I_\nu$ is an energy interval associated with each partial
wave $(l_\nu,j_\nu)$.
The matrix elements of the interaction
involving states in the continuum are given by 
$\tilde{V}_{i\overline{i},\nu\epsilon\overline{\nu\epsilon}} \equiv
\langle \psi_i\psi_{\overline{i}} \vert V
\vert \psi_\nu(\epsilon)\psi_{\overline{\nu}}(\epsilon) \rangle$,
$\tilde{V}_{\nu\epsilon\overline{\nu\epsilon},
\nu^\prime\epsilon^\prime\overline{\nu^\prime\epsilon^\prime}}
\equiv
\langle \psi_\nu(\epsilon)\psi_{\overline{\nu}}(\epsilon) \vert V\vert
\psi_\nu^\prime(\epsilon^\prime)\psi_{\overline{\nu^\prime}}
(\epsilon^\prime)\rangle$. The rest of the notations are standard
\cite{Schuck}. 
It can be noticed that according to the BCS approach 
the generalized gap equations above take into account only pairing 
between time-reversed continuum states $\nu\epsilon,
~\overline{\nu\epsilon}$. A more general pairing between continuum states at 
neighbouring energies is conceivable and this would just be taken care of by a 
continuum HFB approach. At the moment there exist only HFB calculations performed 
with a box boundary condition and we shall numerically compare their results with 
those obtained in the present approach.    

The largest contributions to the integrals above 
come from 
the regions where the wave functions $\psi_\nu(\epsilon)$
have a large localization inside the nucleus.
This condition is fulfilled
in  energy  regions where the $S$-matrix has poles close  to  the
real energy axis,
i.e., near narrow single-particle resonances.
The integrals can also have
large contributions
from energy intervals close to zero energy
if the
$S$-matrix has poles corresponding to loosely bound states
or virtual states near threshold \cite{Migdal,MPP}.
In the $I_{\nu}$ intervals defined above the positive energy wave functions 
have the largest localization inside a sphere of radius $D$, where $D$ 
is of the order of a few times 
the nuclear radius. Within $D$ the 
positive energy wave funtions can be related to the scattering wave function
at resonant energy $\epsilon_\nu$ through
 simple factorization formulas \cite{Migdal,MPP,Unger,Giai}. 
Following Refs.\cite{Unger,Giai},
the wave function $\psi_\nu(\epsilon)$ inside $D$ can be approximated by
 \begin{equation}\label{eq:fact}
 \psi_\nu(\epsilon,r) \approx 
 g^{1/2}_\nu(\epsilon)\tilde{g}^{-1/2}_\nu(\epsilon)
 \phi_\nu(\epsilon_\nu,r)
 \equiv 
  \tau^{1/2}_\nu(\epsilon)\phi_\nu(\epsilon_\nu,r) ~ ,
 \end{equation}
where $\phi_\nu(\epsilon_\nu,r)$ is the scattering wave
function calculated at the resonant energy $\epsilon_\nu$ and
normalized within a sphere of radius D.
 This factorization relation is very useful for evaluating
 matrix elements of finite range interactions 
for which it is sufficient to carry space integrals
over the volume inside the radius $D$ only. For instance, we will use 
 $\tilde{V}_{i\overline{i},\nu\epsilon\overline{\nu\epsilon}} \approx
\tau_\nu(\epsilon)
V_{i\overline{i},\nu\epsilon_\nu\overline{\nu\epsilon_\nu}}$
and  $\tilde{V}_{\nu\epsilon\overline{\nu\epsilon},
\nu'\epsilon'\overline{\nu'\epsilon'}} 
\approx \tau_\nu(\epsilon)\tau_{\nu'}(\epsilon') 
V_{\nu\epsilon_\nu\overline{\nu\epsilon_\nu},
\nu'\epsilon_{\nu'}\overline{\nu'\epsilon_{\nu'}}}$, 
where on the right hand sides the matrix elements of the interaction
are calculated using the wave functions $\phi_\nu(\epsilon_\nu,r)$.
 For a discussion of the accuracy of these approximations see
Ref.\cite{Giai}.

With the help of this factorization  the gap 
equations~(\ref{eq:gap1},\ref{eq:gap2}) become:
\begin{equation}\label{eq:gapr1}
\Delta_i = \sum_{j}V_{i\overline{i}j\overline{j}} u_j v_j +
\sum_\nu
V_{i\overline{i},\nu\epsilon_\nu\overline{\nu\epsilon_\nu}}
\int_{I_\nu} g_{\nu}(\epsilon)
u_\nu(\epsilon) v_\nu(\epsilon) d\epsilon~,
\end{equation}
\begin{equation}\label{eq:gapr}
\Delta_\nu \equiv
\sum_{j}
V_{\nu\epsilon_\nu\overline{\nu\epsilon_\nu},j\overline{j}} u_j v_j +
\sum_{\nu^\prime}
V_{\nu\epsilon_\nu\overline{\nu\epsilon_\nu},
\nu^\prime\epsilon_{\nu^\prime}
\overline{\nu^\prime\epsilon_{\nu^\prime}}}
\int_{I_{\nu^\prime}} g_{\nu^\prime}(\epsilon^\prime)
u_{\nu^\prime}(\epsilon^\prime) v_{\nu^\prime}(\epsilon^\prime) 
d\epsilon^\prime~,
\end{equation}
with $\Delta_\nu(\epsilon) =\tau_\nu(\epsilon)\Delta_\nu$.
The last expression can be written as
$\tilde{g}_\nu(\epsilon)\Delta_\nu(\epsilon)=g_\nu(\epsilon)\Delta_\nu$ 
and gives the connection between the gaps calculated with the
wave functions $\psi_\nu(\epsilon)$ and $\phi_\nu(\epsilon)$.
One can get the same relation if one writes the gap equation (\ref{eq:gap2})
in terms of a local pairing field $\Delta(r)$ of finite range which
cuts off the contributions of the 
tail of the wave function $\psi_\nu(\epsilon,r)$ beyond the 
radius $D$. Thus, 
\begin{equation}\label{eq:cut}
\Delta_\nu(\epsilon)  =  
\int_0^{R_b}|\psi_\nu(\epsilon,r)|^2\Delta(r)dr
\approx 
\tau_\nu(\epsilon)\int_0^{D} |\phi_\nu(\epsilon,r)|^2 \Delta(r)dr
 \equiv  \tau_\nu(\epsilon)\Delta_\nu~.
\end{equation} 
A similar relation can be derived for the positive energy single-particle 
spectrum. By using these relations it can be seen that the gap equations 
(\ref{eq:gapr1},\ref{eq:gapr}) are independent of the box radius.
As shown above, this is a consequence of the finite range of the pairing 
interaction, which 
is sensitive only to 
the inner  part of the resonant continuum wave functions.

 In the BCS approximation the number of particles is fixed
consistently with the gap equations by counting the particles 
distributed in the model space in which the pairing interaction 
is effective. Using the same approximations as for 
deriving the gap equations  one gets
\begin{equation}\label{eq:n}
N = \sum_{i}v_i^2 +
\sum_\nu \int_{I_\nu} {g}_{\nu}(\epsilon) v^2_\nu(\epsilon)d\epsilon~.
\end{equation}
This equation together with the gap equations (\ref{eq:gapr1},\ref{eq:gapr})
are the extended BCS equations 
for a general (finite range) pairing interaction including the contribution 
of the resonant continuum. 
Each resonance is characterized be the quantity $\Delta_\nu$,
which acts as the averaged gap of that resonance. 
  
The above BCS equations are well suited to 
be specialized to 
the approximation of constant pairing interaction since
the resonant states $\phi_{\nu}(\epsilon_\nu,r)$ and the 
bound states have rather similar localizations inside the nucleus.
Then, one just has to replace 
in Eqs.(\ref{eq:gapr1},\ref{eq:gapr}) all matrix elements by a constant 
value $G$. The corresponding BCS equations are the same as 
the ones of Ref.\cite{NS}. It is worthwhile to point out that in the 
constant pairing approximation as defined here one preserves the
variation of the matrix elements of the pairing interaction over the 
resonance region. Consequently, as seen in Eq.(\ref{eq:cut}),
the gap also changes in the resonance region and therefore the 
corresponding pairing field is not constant in the whole space.

We  can  now extend the above treatment of the resonant continuum to HF+BCS.
In the case of a Skyrme force this is done
by including  into the nucleon densities the contributions of the positive energy
states with energies in the selected  intervals $I_{\nu}$ and by using 
the factorization relation (\ref{eq:fact}). Thus the resonant 
continuum contribution 
to the particle density inside the sphere of radius $D$ reads
\begin{equation}\label{eq:dens}
 \rho_c(r) 
  \approx  \sum_\nu\vert \phi_\nu(\epsilon_{\nu},r)\vert ^2
  \int_{I_\nu} g_\nu(\epsilon) v^2_\nu(\epsilon)d\epsilon
 \equiv  
\sum_\nu\vert \phi_\nu(\epsilon_{\nu},r)\vert ^2\langle v^2 \rangle_{\nu}
~.
\end{equation} 
Similar expressions can be derived for the kinetic energy density 
$T(r)$ and spin density $J(r)$,
\begin{equation}\label{eq:kin}
 T(r) \approx
 \sum_{i} v^2_i \vert \nabla \psi_i(r)\vert ^2 +
 \sum_{\nu}
 \langle v^2 \rangle_{\nu} \vert \nabla \phi_\nu(\epsilon_{\nu},r)\vert^2
  ~,
 \end{equation}
\begin{equation}\label{eq:ang}
 J(r) \approx
 -i \sum_{i} v^2_i \psi_i^*(r) (\nabla \psi_i(r)\times \sigma)
 -i \sum_{\nu} \langle v^2 \rangle_{\nu}
 \phi_{\nu}^*(\epsilon_\nu,r)
 (\nabla \phi_{\nu}(\epsilon_\nu,r)\times \sigma)~,
\end{equation}
where the first sum represents the contribution of the bound states. 
The above densities 
define the mean field and the single-particle spectrum. They depend 
on the occupation probabilities and they are calculated iteratively with the 
BCS equations, as in the usual HF+BCS calculations\cite{Vautherin}. 
 
At this point we would like to comment on the relation between the HF+BCS 
 equations derived above and the HFB approach. 
 The advantage of the  HFB approach to
treat  processes that involve the continuum part of the  nuclear 
spectrum is that the finite range of the pairing field is explicitely 
 taken into account. Therefore, the particle and pairing densities 
acquire  automatically a proper asymptotic behaviour 
\cite{Bulgac,Dobaczewski}.  
 In order to preserve the same behaviour in the HF+BCS limit one should keep
 the physical condition of a finite range pairing field, as it is also done in 
 HFB calculations in which the pairing field is not calculated 
 self-consistently\cite{Belyaev}. 
 As seen in Eq.(\ref{eq:cut}), a finite range pairing field implies a cut-off 
 in the tail 
 of the positive energy  wave functions. 
Without this cut-off the solution of a HF+BCS
 calculation with the positive energy states discretized in a box would 
 correspond to a nucleus in dynamical equilibrium with a nucleonic gas and not 
 to the nucleus itself\cite{Bonche}. Generally the cut-off radius may be 
 ambiguous but if one restricts oneself to the resonant 
 continuum, then there is a rather large region outside 
 the nucleus where the resonant wave functions have values 
 close to zero before they start oscillating. In this case 
 the HF+BCS results do not depend sensitively
 on the cut-off radius chosen in such region. 

We now apply the above resonant HF+BCS approach to a nucleus far from stability,  
namely $^{84}$Ni for which HFB results can be found \cite{Terasaki}. Here, 
we wish to compare the results of three types of calculations: A) the resonant 
HF+BCS approach where the widths of single-particle resonances are taken into 
account; B) a discrete version of it where the widths are set to zero; C) the HFB 
approach of Ref.\cite{Terasaki} where the coordinate space equations are solved 
with a box boundary condition and therefore the width effects would be missing.   
The HF field is calculated with the SIII interaction \cite{SIII} whereas 
for the pairing channel 
a zero-range density-dependent interaction is used,  
   $V(\vec{r}_1,\vec{r}_2)= V_0 (1-\rho(\vec{r}_1)/\rho_c) 
   \delta(\vec{r}_1-\vec{r}_2)$, 
 where $\rho(\vec{r})$ is the total density, $V_0=1128.75$ MeV fm$^3$ and 
 $\rho_c=0.134$ fm$^{-3}$. 
All calculations are carried out up to a distance $D$= 22.5 fm (see Eq.(4)) 
but the numerical results discussed
 in the following do not depend sensitively on the precise choice of 
 $D$ in the range $(3 - 4)R$, where $R$ is the nuclear radius. 
 The resonant states included in the HF+BCS calculations, with  energies 
 smaller than 5 MeV (the energy cut off used in Ref.\cite{Terasaki}), 
 together with  the last bound state are listed in Table 1.  
 The energy $\epsilon_\nu$ (width $\Gamma_\nu$ ) of a given resonance 
 is extracted from the energy where the derivative of the 
 phase shift reaches its maximum (half of its maximum) value. The energy 
intervals $I_\nu$ are defined such that 
$\vert \epsilon - \epsilon_\nu \vert  \le 2\Gamma_\nu$.

Let us first look at the results of case A.  
The total averaged gap and the Fermi energy are  
 $<\Delta>$= 0.51 MeV
 and $\lambda$=-0.874 MeV, respectively. The total 
 binding energy and the pairing energy are $E$=-652.7 MeV
 and $E_{P}$=3.4 MeV.
 The corresponding pairing field is shown in Fig. 1 while the averaged 
 occupation probabilities and the averaged gaps of resonant 
 states and the last bound state $3s_{1/2}$ are given in Table 1. 
  The change of the particle density due to 
 pairing correlations is shown in Fig. 2. It can be seen that in the
 tail region the contribution of the bound states to the total density, 
 given mainly by the loosely bound state $3s_{1/2}$, is dominant. 
 In order to see how the neutrons are distributed at large distances, 
 we have calculated the number of neutrons outside a sphere of radius 
 12 fm. We find that the total numbers of neutrons 
 distributed in bound and resonant states between 12 and 22 fm are  
 0.069 and 0.037, respectively. 

 A proper estimation of the particle distribution at large distances
 is difficult in HFB calculations based on a box boundary condition.
 Due to the box the wave functions which are spread far from 
 the nucleus are generally pushed towards smaller distances. 
 Thus, in the HFB calculations of Ref.\cite{Terasaki} the occupancy of 
 the loosely bound state $3s_{1/2}$ which gives the dominant contribution 
 in the density tail region, depends strongly on the box radius and is 
 generally underestimated. Consequentely the HFB density is smaller in 
 the tail region than the HF+BCS density. As seen in Fig. 2 the tail
 of the HFB density is actually smaller even than the HF density.

 For case B the total averaged gap becomes   
 $<\Delta >$= 0.72 MeV and the Fermi energy is $\lambda$= -0.948 MeV. 
 The binding energy increases to the value $E$=-653.1 MeV
 while the pairing energy is equal to $E_{P}$=6.2 MeV. 
 From Table 1 one can see that the occupancy of the resonant states
 is almost doubled compared to the case when the effect of the widths
 is taken into account. 
 The corresponding changes in the 
 pairing field and particle density are shown in Figs. (1,2). 
 As seen in Fig. 2 the pairing field  
 given by the HFB calculations based on a box boundary condition
 are quite similar to the HF+BCS pairing field of case B, i.e., calculated 
 without taking into account the width effect. The same similarities
 are seen in the Fermi, binding and pairing energies, 
 which in the case of HFB are: $\lambda$=-0.956 MeV, 
 $E$=-653.7 MeV and $E_{P}$=6.9 MeV. These similarities show that 
 in the HFB calculations with a box boundary condition the resonant
 continuum is actually described by quasibound states and therefore in
 such calculations the effect of the widths of resonant states upon
 pairing properties is not taken into account properly.
   
  In conclusion one sees that neglecting 
 the contribution of the widths in HF+BCS calculations one enhances 
 artificially the amount of pairing correlations. 
 This enhancement is due to the fact that
 if the width is neglected then all
 the pairing strength is collected from the scattering  state at the
 resonance energy. At this energy the scattering wave function 
 has the highest spatial concentration (as compared with the 
 nearby scattering functions) within  the nuclear region. The effect of the
 width is to diminish the pairing strength because the pairs can now
 scatter also in the nearby states around the resonance energy which are less
 confined inside the nucleus. In a time dependent picture the 
 dependence on the width seen above would  
 correspond to the fact that a pair scattered to a resonance state has a
 finite lifetime, thus contributing less to the pairing correlations
 as compared to the case of a very narrow (quasibound) state.
 In the HF+BCS approach presented here this effect is taken into account
 automatically through the continuum level density.

 Formally, in a
coordinate space HFB approach the contribution of the whole
 continuum is taken into account and therefore the effect of the
 resonant continuum discussed above should be also present.
  Furthermore, a HFB
 calculation contains also correlations from pairs in states which
 are not time-reversal partners.
 These correlations which are absent in a HF+BCS
 approach could be important for the  scattering states with
 energies close to a resonance because their wave functions 
 have similar localization properties \cite{Bennett}.
   However, the estimation of the effect of the resonance widths
 upon pairing correlations is still an open problem in the existing
 self-consistent HFB calculations. This is because the numerical methods
 used for solving the coordinate space HFB equations are based on
 discretizations of the continuous spectrum. With the currently used values
 of box radius (R $\simeq$ 15-30 fm) each resonance is represented by a
 single discrete state in the spherically symmetric case.  
To have a discrete level density high enough to describe properly the shape
of the resonance would require an extremely large box radius, a condition
which makes practical calculations untractable. 
In addition, the use of a 
 box of finite radius could alter significantly the description of the 
 pairing correlations induced by the loosely bound states whose tail might 
 extend beyond the box radius. 
 Although the 
coordinate space HFB approach is in principle the appropriate tool for 
 treating continuum effects, solving the HFB equations by imposing a box
 boundary condition to the solutions 
 does not guarantee that all effects of the continuum, 
 particulary the effect of the widths of resonant states, are 
 properly taken into account.

  In summary, a method to include the resonant continuum in the
  HF+BCS approximation is presented. We have here concentrated on the
  regions of the continuous spectra which are close to single-particle
  resonances because they bring the most important contributions to 
  pairing correlations. 
  The method can be used to take also into account 
  the effect of non-resonant continuum states close to the continuum
  threshold, which can be important in the presence of loosely bound 
  states or virtual states. 
  In the numerical example it has been shown that the widths 
  of resonant states have an important effect on the pairing properties 
  of nuclei close to the drip line. In order to describe this 
  effect in a self-consistent HFB approach one needs to solve the
  HFB equations with proper boundary conditions. This work is in 
  progress. 

  We thank P.H.Heenen and J. Terasaki 
  for useful discussions and for the detailed results of their  
  HFB calculations. 
  Two of us (NVG and NS) would like to thank the IN2P3-INPE collaboration
  in the frame of which part of this work was performed. NS acknowledges 
  the financial support of the Wenner-Gren Foundation.

\begin{table}
\label{table1}
\caption{
Results of HF+BCS calculations for the nucleus $^{84}$Ni.
$\Delta_n$ and $v^2_n$ are the averaged gap and averaged
occupation  probability of the single-particle state $n$ of energy
$\epsilon_n$ and width $\Gamma_n$. The notations $\tilde{\Delta}_n$,
$\tilde{v}^2_n$, $\tilde{\epsilon}_n$ and $\tilde{\Gamma}_n$ stand
for the corresponding quantities calculated without including the
effect of the widths of resonant states in HF+BCS equations.
The single-particle energies, their widths and the pairing gaps are 
expressed in MeV. }
\vskip 0.6cm
$\begin{array}{||cc|cc|cc|cc|cc|cc|cc|cc|c||}\hline\hline
n & & \epsilon_n & & \tilde{\epsilon}_n & & \Gamma_n & & 
\tilde{\Gamma}_n & & v^2_n & & \tilde{v}^2_n & & \Delta_n & &
 \tilde{\Delta}_n \\\hline\hline
s_{1/2} & & -0.647 & & -0.644 & & ---  & &  --- & & 0.295 & &
 0.294 & & 0.505 & & 0.674 \\
d_{3/2}  & &  0.441 & &  0.417 & & 0.077 & & 0.068 & & 0.041 & &
 0.075 & & 0.630& & 0.847 \\
g_{7/2}  & &  1.604 & &  1.587 & & 0.009 & & 0.008 & & 0.029 & &
0.055 & & 0.966& & 1.306 \\
h_{11/2} & &  3.309 & &  3.302 & & 0.017 & & 0.016 & & 0.017 & &
 0.034 & & 1.227 & & 1.658 \\\hline\hline
\end{array}$
\end{table}

\vskip 2cm

\centerline{ FIGURE CAPTIONS}
\vskip 1cm

Figure 1: Neutron pairing field as a function of the radius.
 The full (dashed) line shows the results of the HF+BCS calculations
 with (without) the effect of the width included. The line marked
 by crosses shows the HFB results of Ref.\cite{Terasaki}. 

\vskip 0.5cm

Figure 2: Neutron density in $^{84}$Ni, calculated in HF (long dashed line)
 and HF+BCS. In the order of decreasing tail the results of HF+BCS
 correspond to the following approximations: effect of the
 widths neglected; effect of the widths included;
 contribution of the bound states to the density.
 All densities are in fm$^{-3}$.
 The inset shows the corresponding densities in linear scale.

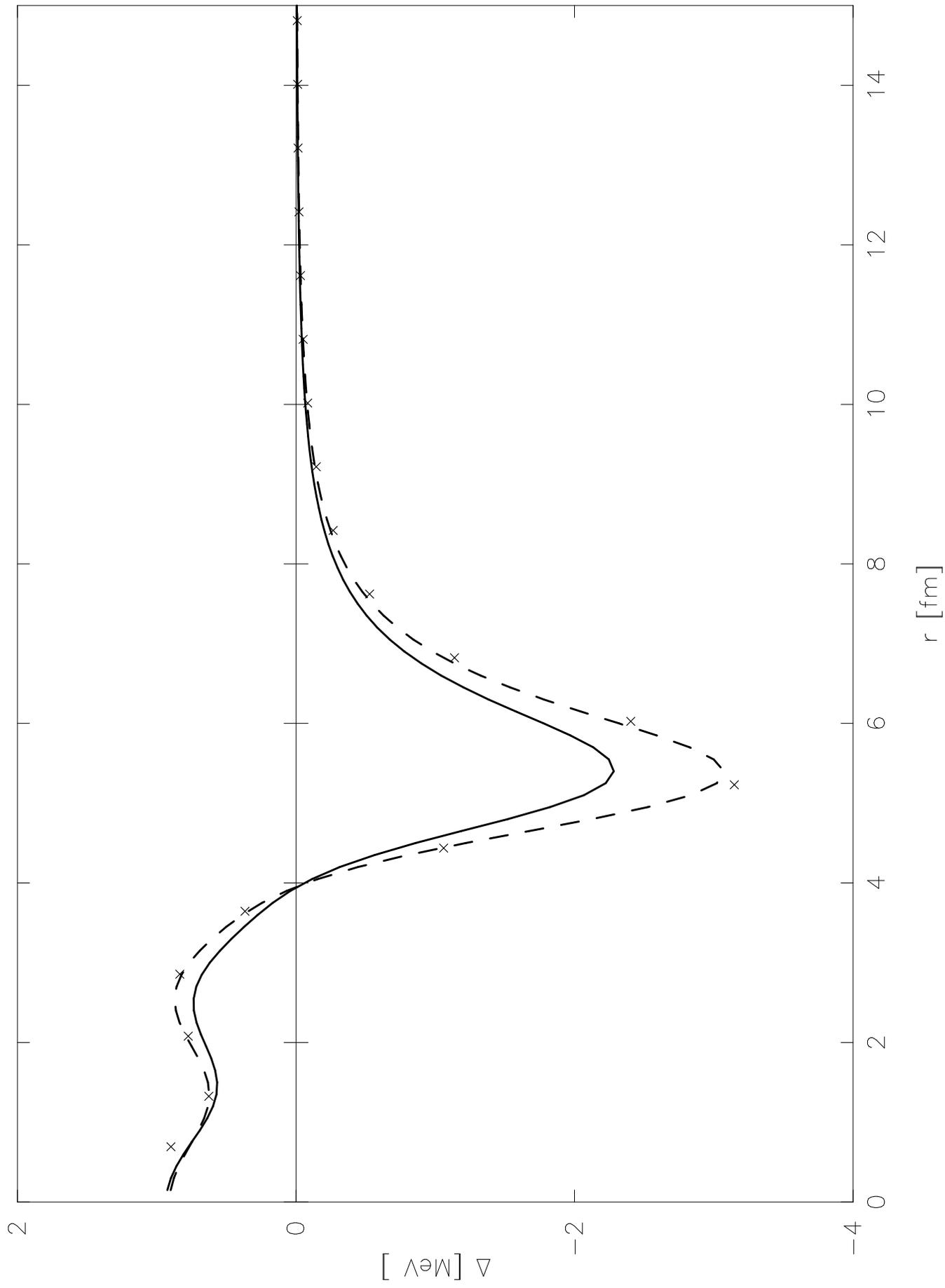
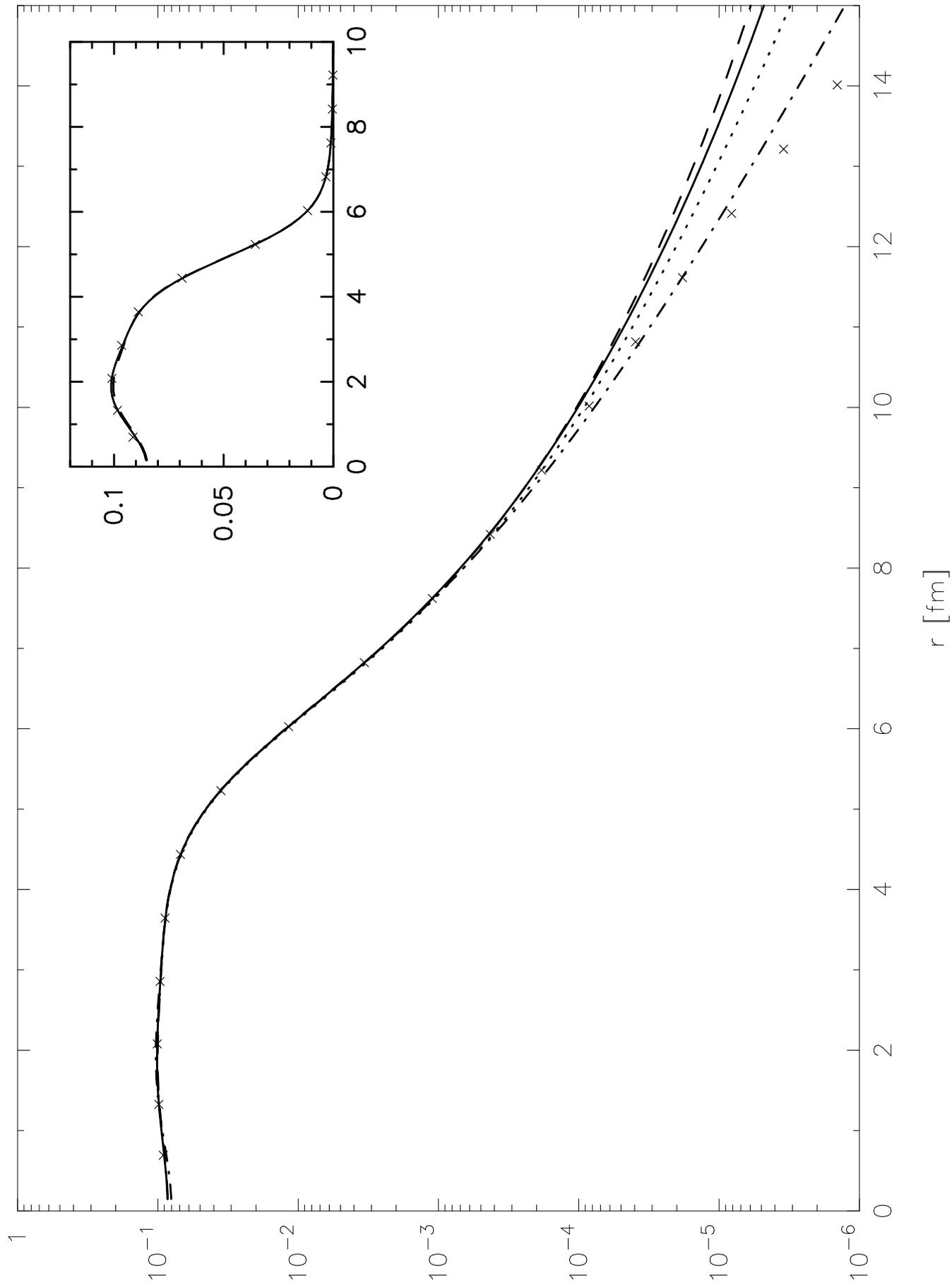
\end{document}